\documentclass[10pt, conference, compsocconf]{IEEEtran}
\usepackage{cite}
\ifCLASSINFOpdf
  \usepackage[pdftex]{graphicx}
\else
\fi

\usepackage{graphicx}
\usepackage{booktabs} % For formal tables
\usepackage{fancyhdr}

\usepackage{amsthm}
\usepackage{amsmath}
\usepackage{caption}
\usepackage{algpseudocode}
\usepackage{epstopdf}
\usepackage{subfig}

\usepackage{multirow}
\usepackage{colortbl}
\usepackage[english]{babel}
\usepackage[ruled,vlined,linesnumbered,nofillcomment]{algorithm2e}
\usepackage[flushleft]{threeparttable}

\newcommand\circledblack[1]{
  \ooalign{%
    \hidewidth
    \kern0.8ex\raisebox{-1.2ex}{\scalebox{3.5}{\textcolor{black}{\textbullet}}}
    \hidewidth\cr
    $\textcolor[rgb]{1.00,1.00,1.00}{\bf {#1}}$\cr
  }%
}

\begin{document}
\title{Scaling Up Large-Scale Graph Processing for GPU-Accelerated Heterogeneous Systems}

\author{\IEEEauthorblockN{Xianliang Li}
\IEEEauthorblockA{Huazhong University of Science and Technology\\
xianliang@hust.edu.cn}
}

\maketitle

\begin{abstract}
Not only with the large host memory for supporting large scale graph processing, GPU-accelerated heterogeneous architecture can also provide a great potential for high-performance computing. However, few existing heterogeneous systems can exploit both hardware advantages to enable the scale-up performance for graph processing due to the limited CPU-GPU transmission efficiency.

In this paper, we investigate the transmission inefficiency problem of heterogeneous graph systems. Our key insight is that the transmission efficiency for heterogeneous graph processing can be greatly improved by simply iterating each subgraph multiple times (rather than only once in prior work) in the GPU, further enabling to obtain the improvable efficiency of heterogeneous graph systems by enhancing GPU processing capability. We therefore present Seraph, with the highlights of {\em pipelined} subgraph iterations and {\em predictive} vertex updating, to cooperatively maximize the effective computations of GPU on graph processing. Our evaluation on a wide variety of large graph datasets shows that Seraph outperforms state-of-the-art heterogeneous graph systems by 5.42x (vs. Graphie) and 3.05x (vs. Garaph). Further, Seraph can be significantly scaled up over Graphie as fed with more computing power for large-scale graph processing.
\end{abstract}

\IEEEpeerreviewmaketitle

\section{Introduction}
Graph is a well-known data structure that can represent a wealth of relationship between objects. Graph processing has a great potential to solve many real-world problems, e.g. path navigation, social network analysis and financial fraud detection. As the graph size is increasingly growing up, it has become a critical turning point regarding how to store as well as further process these large-scale graphs efficiently.

A typical solution for large-scale graph processing is to divide the entire graph into many sub-graphs that are then distributed onto different machines for the computation~\cite{malewicz2010pregel,gonzalez2014graphx}. Though these distributed systems (with more computation resources and storage resources) have made the impressive progress~\cite{gonzalez2012powergraph,zhu2016gemini}, people are often more inclined to use a single-machine processing system, which is easier to manage and understand~\cite{kyrola2012graphchi}.

A wide spectrum of graph systems have emerged for processing large-scale graphs under a single machine\cite{shun2013ligra,nguyen2013lightweight}, particularly in the aspect of GPU acceleration because of its powerful computing capacity~\cite{wang2016gunrock,zhong2014medusa,khorasani2014cusha}. With an elegant advance-filter programming model, Gunrock~\cite{wang2016gunrock} naturally integrates many well-optimized graphics analysis techniques. However, neither of these graph systems are able to handle the large-scale graphs that can not fit into the GPU global memory.

In an effort to cope with this problem, researchers extend to store the graph data in the large host memory for assisting the GPU computing. With the vertex data stored in the GPU memory, GTS~\cite{kim2016gts} streams the subgraph data in an asynchronous manner. GraphReduce~\cite{sengupta2015graphreduce} only transfers the subgraph that has at least one active vertex or edge to the GPU. Unfortunately, due to relatively-low interconnect bandwidth between host and GPU (e.g., $\sim$12GB/s for PCI-Express 3.0), the potential limitation of GPU accelerator under the heterogeneous architecture can become more serious.

Our motivating study (discussed in Section 2.2) also shows that, 75\% of GPU computing capability can be under-utilized even in the presence of existing state-of-the-art GPU-specific optimizations. It gradually becomes of great importance and necessity to scale up the performance of heterogeneous graph systems for large-scale graph processing. In this paper, we focus on studying whether and how we can provide a scale-up efficiency of heterogeneous graph systems under a commodity heterogeneous architecture.

Recently, there still exist a number of graph systems that attempt to improve the performance of heterogeneous graph systems. Graphie~\cite{han2017graphie} proposes two renaming algorithms to improve the memory access efficiency, and keeps track of the active partitions to avoid moving the inactive partitions to GPU. Garaph~\cite{ma2017garaph} reduces the transmission amount by performing a part of sparse vertex updating on the host side. Nevertheless, for many real-world large graphs that can be dense and always active, these recent advances may still involve non-trivial amount of data transmission, making them limited for practical use.

In this paper, we present Seraph, a novel heterogeneous graph system, which can significantly scale up the performance of out-of-GPU-memory graph processing. The key insight of this work lies in a fact that each subgraph in one transmission iteration for many graph algorithms (e.g., BFS and SSSP) may involve much information that is useful for the convergence of the next subgraph iteration. Unlike most of existing research that process each subgraph only once and then overwrite their updates~\cite{han2017graphie, ma2017garaph}, we propose to iterate each subgraph multiple times so as to fully exhaust the value of each subgraph for avoiding the redundant data transmission and unnecessary iteration. Guided by this principle, we can thus make an innovation of leveraging the powerful yet `limited' GPU processing capability to accelerate the multi-time subgraph iteration, enabling the scale-up efficiency for heterogeneous graph systems.

In an effort to break the update limitation within the subgraph, we present to pipeline the iteration of subgraphs. Compared with CLIP~\cite{ai2017squeezing} that iterates on a fixed subgraph loaded from disk, our pipelined subgraph iteration is novel with a maximum scope of the subgraph information propagation.
Further, based on existing high-optimized GPU-based computing model (e.g., pull execution model), we obverse that a large number of vertex computations do not necessarily contribute to a valid vertex updating, causing to waste a large amount of GPU processing capacity. We propose a predictive vertex updating, aiming at efficiently identifying these vertices and further eliminating the unnecessary computations on them for better supporting pipelined subgraph iteration.

We compare Seraph with two state-of-the-art heterogeneous graph systems. Our results on a wide variety of real-world graphs demonstrate that Seraph outperforms Graphie~\cite{han2017graphie} and Garaph~\cite{ma2017garaph} by 5.42x and 3.05x, respectively. In particular, Seraph can be significantly scaled up over existing heterogeneous graph systems. In addition, we compare Seraph with other large-scale graph processing solutions, revealing that Seraph can also achieve impressive performance in comparison to state-of-the-art CPU-based (i.e., Ligra~\cite{shun2013ligra}) and distributed graph processing systems (i.e., Gemini\cite{zhu2016gemini}).

The rest of this paper is as follow. Section 2 gives the background and motivation. We present pipelined iteration in Section 3. Section 4 elaborates predictive vertex updating. Section 5 shows the results. We survey the related work in Section 6. Section 7 concludes the work.
\begin{figure}
\includegraphics[scale=0.6]{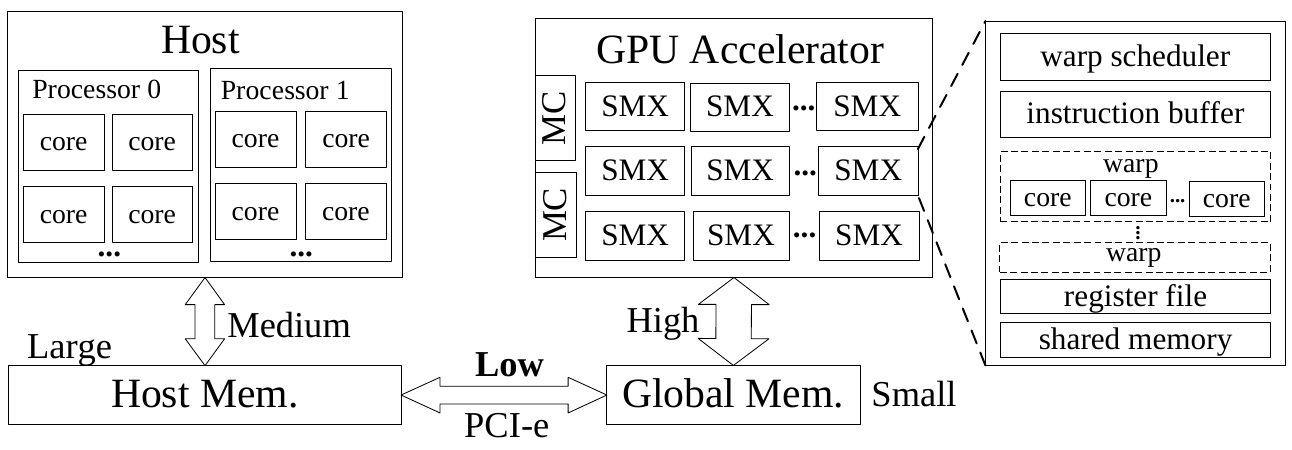}
\vspace{-1em}
\caption{GPU-accelerated Heterogeneous Architecture}
%\vspace{-1em}
\label{fig:Heterogeneous Architecture}
\end{figure}

\section{Background and Motivation}
In this section, we first give a brief introduction to GPU-accelerated heterogeneous architecture, followed by a motivating study regarding the inefficient data transmission of existing heterogeneous graph systems for large-scale graph processing, finally motivating our approach.

\subsection{Heterogeneous Architecture}
There emerge various heterogeneous architectural designs. Some are dedicated to performance improvement \cite{kayiran2014managing}, some for energy reduction \cite{wang2014co}. Some take both into consideration \cite{munger2016carrizo}. This paper has focused on GPU-accelerated heterogeneous architecture.

Figure~\ref{fig:Heterogeneous Architecture} illustrates a typical GPU-accelerated heterogeneous architecture, which integrates the hardware advantages of both host side (with larger host memory) and GPU accelerator (with stronger computing ability). A GPU accelerator generally consists of multiple streaming multiprocessors (SMXs), each of which includes hundreds of cores. 

In comparison to the high-speed internal bandwidth (e.g., $\sim$700GB/s for NVIDIA Tesla P100) of GPU cores accessing global memory, GPUs are generally connected to the host side with the relatively slow interface. For instance, the transmission bandwidth between CPU and GPU via PCI Express 3.0 lane connection can be limited as slowly as $\sim$12GB/s in practice~\cite{ben2017groute}. This significant gap may severely suppress the performance potential of heterogeneous architecture if the data is frequently transferred~\cite{kim2016gts,han2017graphie}.

Though there are a number of transmission interfaces that can provide higher interconnect bandwidth (e.g., Intel QuickPath Interconnect with 25.6GB/s, NVLink high-speed interconnect with 160GB/s), this paper focuses on PIC-e interconnect since it is more common in the current commodity market.
\begin{figure}[t]
\includegraphics[scale=0.55]{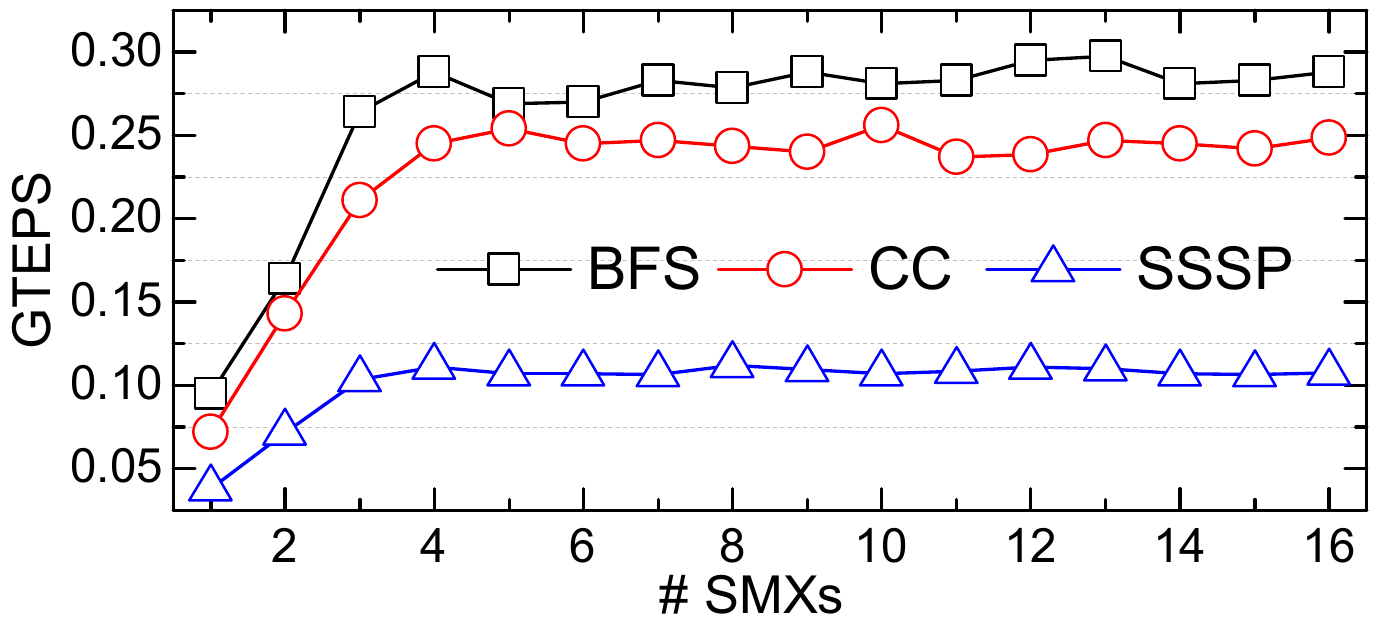}
\vspace{-1em}
\caption{Performance characterization using Graphie's subgraph iteration~\cite{han2017graphie} with the varying GPU SMXs}
\vspace{-1em}
\label{fig:motivation}
\end{figure}
\subsection{Inefficiency of Existing Heterogeneous Graph Systems: A Motivating Study}
In an effort to leverage the hardware advantages of heterogeneous architecture for large-scale graph processing, existing heterogeneous graph systems generally divide the entire graph data into subgraphs~\cite{han2017graphie, ma2017garaph,ai2017squeezing}. The CPU offers the graph data to the GPU in the form of subgraph. %Once each subgraph is consumed, the CPU then sends the next subgraph until the graph algorithm is converged.
Once each subgraph is consumed, the GPU requests to process the next subgraph which is transferred from the host memory to the GPU global memory. %and GPU runtime provide asynchronous stream to overlap the data transmission and kernel execution.

As a representative of state-of-the-art heterogeneous graph processing system, on the basis of basic subgraph iteration above, Graphie~\cite{han2017graphie} additionally has two highlighted optimizations. First, it is ensured that only those subgraphs that have at least one active edge or vertex can be transferred. Second, an asynchronous runtime to reuse the transferred subgraph at the next iteration. As a consequence, Graphie can partly reduce an amount of data transmission. Nevertheless, the potential of GPU-accelerated heterogeneous architecture can be still limited since their approach may be still inefficient to handle many graph algorithms (e.g., CC) where almost all subgraph can be active in the first few iterations. As a consequence, the majority of subgraphs still need to be transferred to the GPU within the iteration.

We investigate the performance characterization of Graphie's cached subgraph iteration on three well-known graph algorithms as the number of available GPU SMXs is increasing in Figure~\ref{fig:motivation}. More details regarding the experimental settings can be found at Section~\ref{sec:evaluation:setup}.Offering more SMXs is helpless for improving the efficiency of large-scale graph processing. As is known, current mainstream GPU accelerators usually {have far more than 4 SMXs}. There remains a significant gap between the low data transmission efficiency and high GPU computing capability for large-scale graph processing. Unfortunately,  existing heterogeneous graph systems rarely respond to this challenge easily for providing the scale-up efficiency.

Under the premise of a fixed transmission bandwidth, one viable approach to improving the CPU-GPU transmission efficiency is to increase the bandwidth utilization with the well-organized subgraph data. Nevertheless, there still remain two significant defects at least for this approach. First, on account of the random memory accesses, graph processing usually behaves poor data locality~\cite{sengupta2015graphreduce, maass2017mosaic}. It is extremely difficult, if not impossible, to prepare a high-quality subgraph that can be fully used for each iteration. Second, even if we can identify such high-quality subgraph, it is also difficult to gather these data in a cost-efficient manner at runtime since graph re-organization is a well-known time-consuming process that may take more time than graph processing for many graph datasets~\cite{zhu2015gridgraph, malicevic2017everything}.

\subsection{Our Observations}
This work aims at reducing the impact of slow subgraph transmission on the high-performance graph computing. Instead of expensively preparing the high-quality subgraphs, we have the key insight that the transmission efficiency of heterogeneous graph systems can be greatly improved by making full use of the value of each subgraph, backed up by our observations as follows.

\underline{\bf\em Observation $1$}:\quad {\em Each subgraph in one iteration has much useful information that serves to the subgraph processing in the next iteration.}
%We note that the information spreading limited by the one-hop sementic, more specific, the vertex information could notify forward the neighborhood vertices but has no ability to inform the neighborhood's neighborhood. And we also note information could be spread more quickly during one iteration to break the on-hop limitation in this iteration.

Each subgraph is structured with many vertices that are associated via the edges. It has been observed that many graph algorithms (e.g., BFS and SSSP) are incremental iteration method that can greatly benefit from iterating each subgraph multiple times~\cite{ai2017squeezing}. More specifically, in one iteration, the information of a given vertex can be only propagated to its neighboring vertices, which still fall short in informing all the vertices that may be involved for graph processing. In order to handle all these vertices, there may need more iteration times to do so, causing to repeatedly load a large amount of redundant graph data. This inspires us of multi-time subgraph iteration to fully exploit its potential value for improving the transmission efficiency.

%Iterating each subgraph multiple times can greatly
%{\bf explaining the above observation}

\underline{\bf\em Observation $2$}:\quad {\em Multi-time subgraph iteration enables to improve the efficiency of heterogeneous graph processing by exploiting GPU processing capability.}

Multi-time subgraph iteration enables the propagation and sharing of information among the subgraphs with multiple hops. This also provides a possibility to reduce the original task with multiple iterations into one iteration. Thus, a new question is how to efficiently iterate each subgraph multiple times, which largely depends on the GPU processing capability. That is, we are allowed to further improve the performance of heterogeneous graph systems through exploiting GPU processing capability.

With the above-discussed observation, the transmission inefficiency problem of heterogeneous graph systems is transferred into a problem of enhancing GPU-based graph processing, for which, a wide of highly optimized techniques can be directly leveraged~\cite{wang2016gunrock}.

\section{Pipelined SubGraph Iteration}
Guided by Observation $1$, we present a pipelined multi-time subgraph iteration, which is designed to make sure the information of each subgraph iteration can be propagated to a larger scope so that the value of each subgraph can be totally exhausted.

\subsection{Preparation}
We start by introducing the requisite preliminaries for the pipelined subgraph iteration. %And figure \ref{fig:arch} give an overview of Seraph.

{\bf Subgraph Organization} Unlike the edge list organization used in prior work~\cite{han2017graphie}, Seraph uses more compact graph structure that manages to minimize the transmission data demand as much as possible. Specifically, Seraph only needs to transfer the Compressed Sparse Column (CSC) to the GPU. In order to process large-scale graph, CSC structure will be cut into much smaller CSC pages which include the a set of continuous vertices with the corresponding incoming edges. Compared to the page structure in~\cite{ma2017garaph}, our data structure is more compact for omitting the destination vertex index array.
%{\em Subgraph Size ....}

{\bf Heterogeneous Execution Model}
Seraph processes large-scale graphs with a heterogeneous execution model. Similar with prior work~\cite{shun2013ligra,zhu2016gemini,liu2015enterprise,ma2017garaph}, Seraph also adopts a density-aware model, which is efficient with less time-
consumption. Our heterogeneous execution can be described as follow. 1) {In the sparse stage}, we compress the active frontier and perform the sparse updating on the CPU with push-based execution model, so as to make sure GPU focuses on the heavy subgraph iteration. 2) {In the dense stage}, we use the pull-based execution model without data contention overhead, and focus on break the transmission bound and improve the computing efficiency.

\subsection{Pipelined Subgraph Scheduling}
The key idea behind our multi-time subgraph iteration is to pipeline the subgraph scheduling for the purpose of maximizing the information propagation of each subgraph to other subgraphs.
\begin{figure}[t]
\centering
\includegraphics[scale=0.35]{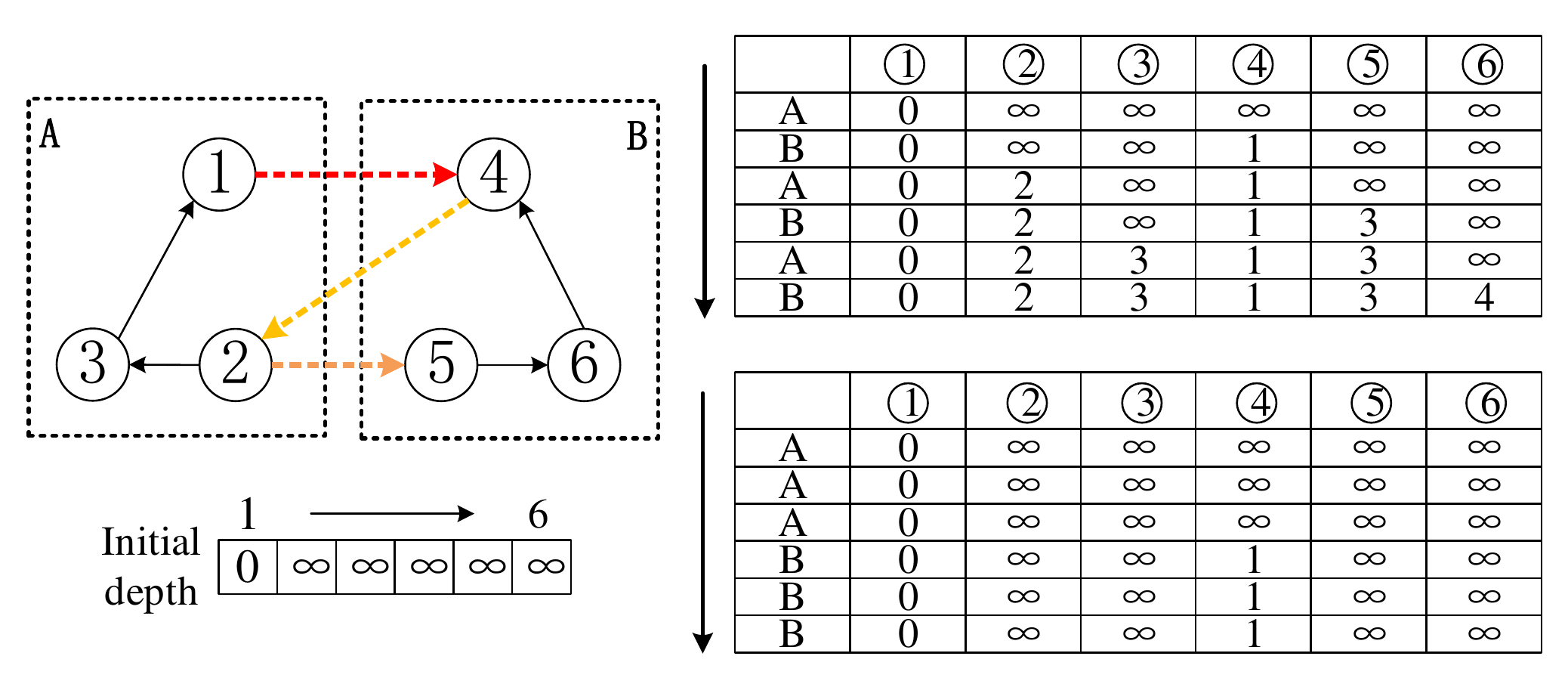}
\vspace{-1em}
\caption{Limited information propagation of subgraph iteration with loaded data reentry in CLIP~\cite{ai2017squeezing}}
%\vspace{-1em}
\label{fig:reentry}
\end{figure}
 Streaming topology is widely used in network optimization~\cite{anyseexfliao}, and also in storage hierarchy optimization. CLIP~\cite{ai2017squeezing} presents a disk-based "loaded data reentry ($ldr$)" streaming topology to squeeze out the value of loaded data. It has the following features: 1) process the subgraph more than once to reduce the total iteration times; and 2) subgraph data is loaded by sequential order to maximize disk IO bandwidth. Nevertheless, the loaded data reentry this way is prone to causing much redundant computation. The underlying reason accounting for this is because the information can not propagate among multiple subgraphs for timely interaction,and block the convergence of the entire graph.

Figure~\ref{fig:reentry} shows an example of CLIP's subgraph iteration with BFS on the given graph based on pull model. Considering the similarity, we only give a part of the entire graph. With the loaded data reentry optimization, we can find that BFS has no sign of convergence, but simply process subgraph one by one has converged quickly with the same schedule times. It is because the subgraph $A$ and subgraph $B$ can not share the update within the multi-reentries. More specifically, the update of $A$ can notify $B$ after $A$ has finished ($1\rightarrow$$4$), but the update of $B$ can not feed back to subgraph $A$ within the multi-reentry ($4\rightarrow$$2$), since subgraph $A$ has been discarded from this iteration. That is, the graph partition destroys the sub-structure among subgraphs, further blocking information propagation between multiple subgraphs.

%Compared with vertex updating cross different subgraphs, the vertex updating within the same subgraph can been viewed immediately, but information propagation is limited within the scope of such subgraph,
%. Unfortunately, graph partition may break down the feedback structure among subgraphs. Thus, we can observe that, supposing the subgraph gradually becomes large, the disk-based "loaded data reentry" method would be degraded into a shared-memory graph processing, without the propagation limits. If we could combine two subgraphs into one subgraph, the information propagation limit among the original two subgraphs should be removed. This inspires us to reentry the subgraph within a larger scope to rebuild the feedback structure. \\

%In this paper, we argue that the useless update is mainly due to the limited graph scope of subgraph.
%Towards this problem, we therefore propose to {\bf xxxx}.
%The main idea of our pipelined is to {\bf xxxx}. {\bf describe the workflow of pipelined subgraph iteration}.
%We observe that the general subgraph-based graph processing has the whole graph as information propagating scope. Inspired by this, we designed a hybrid pipelined subgraph schedule method to gain the benefit of both: with a larger graph propagating scope and squeeze out the value of loaded data .\\
\begin{figure}
\label{fig:sub:schedule}
\includegraphics[scale=0.3]{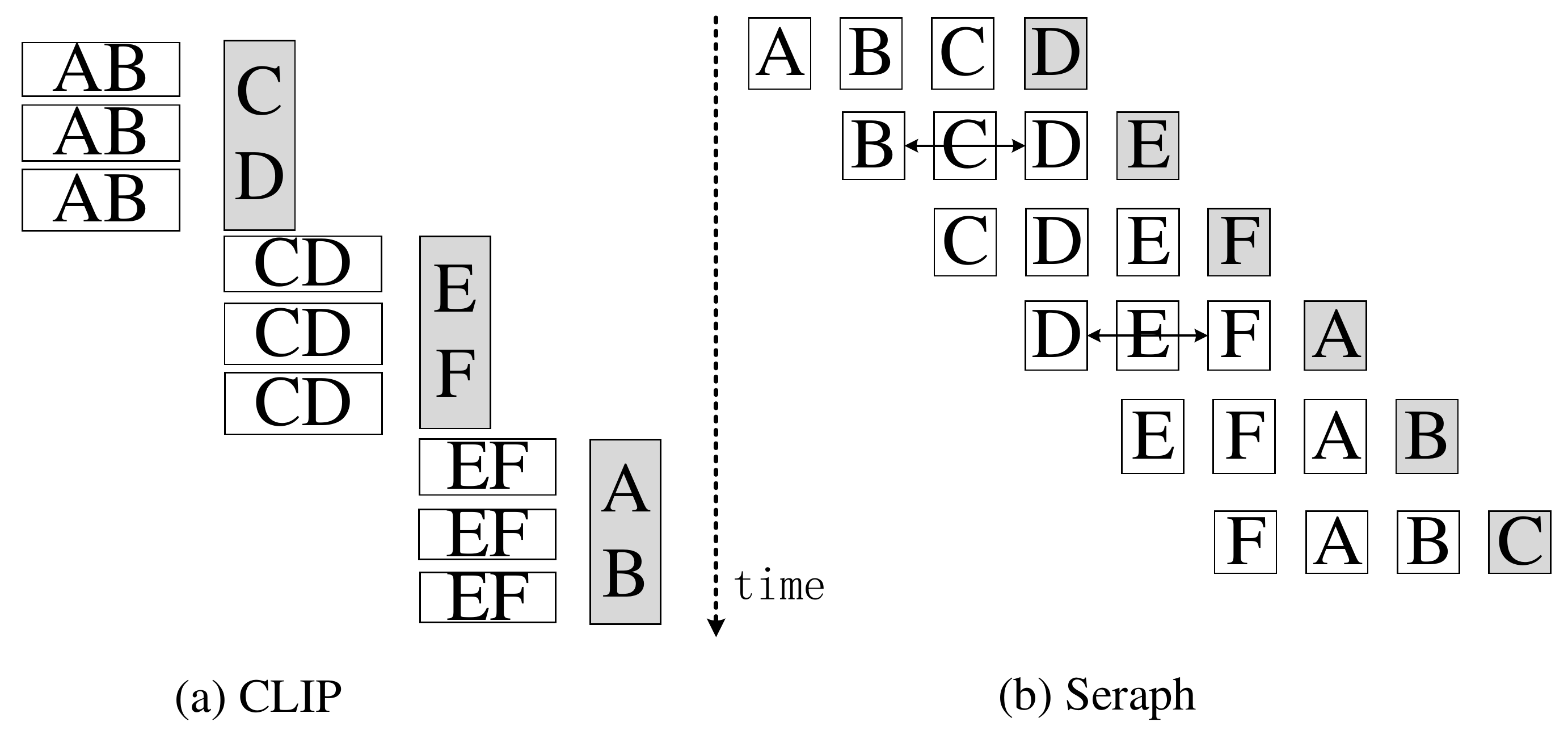}
\vspace{-2em}
\caption{Comparison of the workflow of CLIP's loaded data reentry and Seraph's pipelined subgraph iteration}
\vspace{-1em}
\label{fig:sub:schedule}
\end{figure}
{\bf Pipelined Subgraph Iteration } We therefore present a novel subgraph iteration in a pipelined fashion for maximizing the propagation scope and rebuild the substructure among subgraphs. Our pipelined subgraph iteration method is shown in Figure~\ref{fig:sub:schedule}. For facilitating the descriptions, considering GPU memory space can maximally hold four subgraphs, and we assume that the GPU can process three subgraphs and transfer one subgraph at the same time. Then kernel execution on subgraph A,B,C can be overlapped with the transmission of subgraph D. We could treat subgraph subgraph A, B and C as a ``super subgraph" since we process the vertices and edges in A, B, C concurrently. When the kernel finished, subgraph D has just been copied from the CPU, we then launch kernel to process subgraph B, C, D and transfer subgraph E repeatedly. Afterwards, the space occupied by subgraph A is overwritten.

However, to maximize to substructure and fully overlap the data transfer with kernel execution, double-buffer based schedule workflow can be viewed in figure~\ref{fig:sub:schedule}. Similarly, this method also load the graph once but process more than once, furthermore process the subgraph $AB$ three times and transfer subgraph $CD$ at the same time.

But our approach is more efficient for two reasons. First, we can offer a larger scope for information propagating. As figure~\ref{fig:sub:schedule} shows, the D can even reconstruct substructure with B and E, enlarging the information propagate scope to at most five subgraphs in the example compared with two in $ldr$ used in CLIP~\cite{ai2017squeezing}. Second, pipelined subgraph iteration further allow us to make a fine-grained scheduling, and we introduce as follows.

Example mentioned above assumed that GPU can process three subgraphs and transfer one subgraph at the same time. However, it is hard to reflect most of real-world situations since the execution time is constantly change, and there are many idle slots (transmission wait for computation or computation wait for transmission). In order to fill the idle slots between subgraph transmission and computation, Seraph minimizes the effect by a more fine-grained scheduling. Specifically, if the transmission thread finds that the transmission task has finished, but the computation has not finished yet, it will check if the kernel executed on the first few subgraph (the subgraphs will be discarded from the execution set immediately) has finished. If so, the thread will transfer the next subgraph to GPU, and overwrite the space occupied by the finished kernel. On the contrary, if the computation has finished but the transmission not, Seraph will select one subgraph in the finished kernel set to reentry once. This method mainly benefit from much smaller subgraphs used in Seraph, and offer an incremental model to fill the idle slot as much as possible.
%We note the rest global memory as $G$ (exclude the memory space occupied by the vertices value and other necessary data). Then the maximal scope of subgraph can be $G$, however, this method give no possibility for transferring the next subgraph to overlap the kernel execution. As figure\ref{fig:sub:schedule} shows, a double-buffer based should be more efficient for overlapping the data movement and kernel execution with a relative large subgraph scope($G$/2). But our pipelined subgraph iteration method could offer close to 2$G$ even exceed the maximal memory size with the same iteration times.\\
%{\bf Dynamically Adjusting Subgraph Scope}
%The pipelined sugraph iteration method offers a more flexible schedule opportunity. Seraph enables to adjust a huge subgraph into several smaller subgraphs and rebuild the "large subgraph" within the pipeline. One ideal situation is the graph processing on computation window could perfectly overlapped with the transmission window, to fully match the power of bandwidth and computation. However, it is common that there are a difference between computation and data movement.
%Example mentioned above assumed that GPU can process three subgraphs and transfer one subgraph at the same time. However, it is unrealistic in most cases since the execution time is constantly change. Pipelined subgraph iteration allows the execution set size and transmission set size changing(e.g.,launch kernel on four subgraphs with transferring one subgraph, but launch kernel on subgraphs with ) in the iteration.

{\bf Remark}\quad With subgraph iteration pipelined, we enable to use more information within each iteration for the fewer iteration numbers and faster convergence speed. The vertex update in one subgraph could spread to a large scope rebuilt by multiple smaller subgraph, making the information spreading more sufficient.
\begin{figure}
\centering
\includegraphics[scale=0.3]{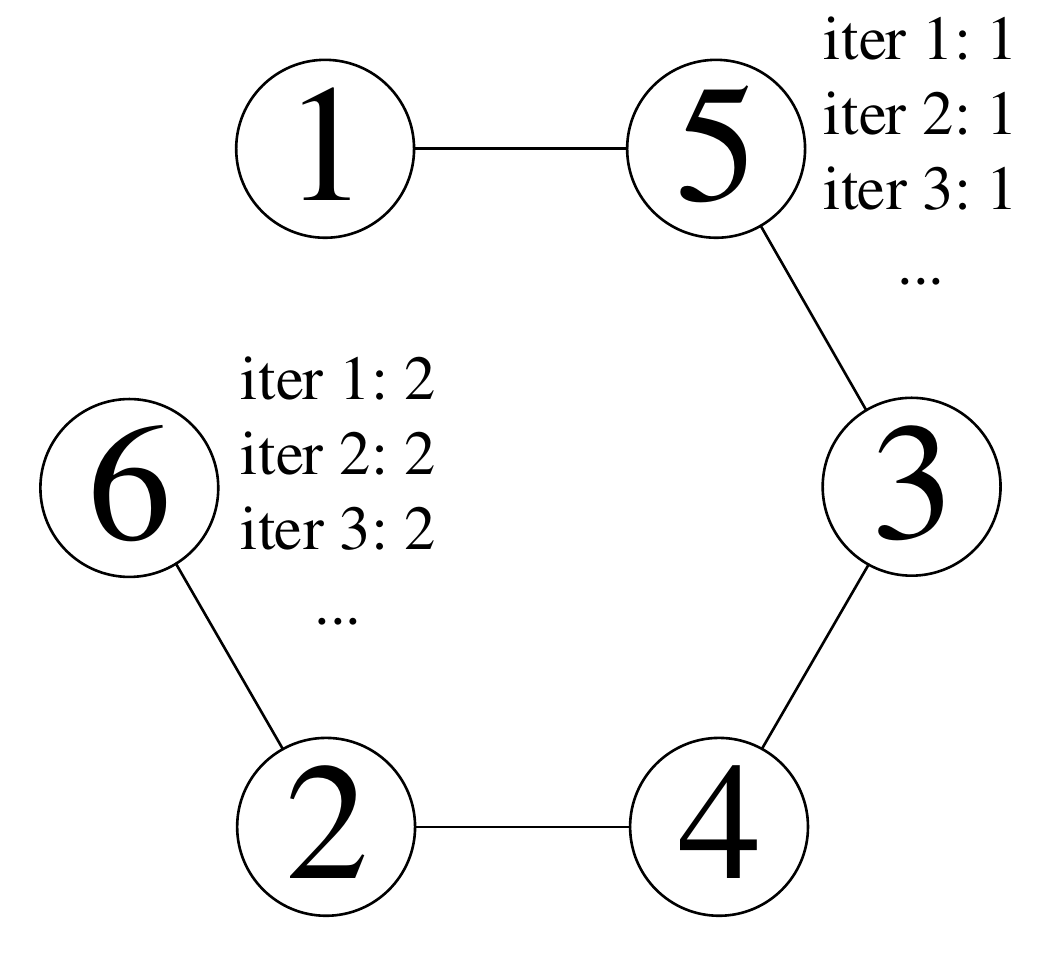}
\vspace{-1em}
\caption{A connected component algorithm on the given graph. We list the labels of vertex 2 and vertex 6 after each iteration}
\vspace{-1em}
\label{fig:pvu:cc}
\end{figure}
\section{Predictive Vertex Updating}
Guided by Observation 2, this section mainly presents predictive vertex updating. We first review the characteristics of vertex updating with pipelined subgraph iteration. In accordance with characteristics analysis, we further present two efficient solutions of vertex updating to enhance the GPU processing capability.

\subsection{Characteristics Analysis of Vertex Centric Updating}
In this work, we follow the pull-based execution model on GPU which involves few data races, so that the data-parallel potential of GPU can be fully exploited.However, almost all existing systems~\cite{shun2013ligra, wang2016gunrock} with density-aware optimization perform the pull update attempt on all vertices. Nevertheless, though a part of vertices can be updated, we note that many of other vertices updating attempts have failed, leading to a large number of unnecessary computations.

Figure~\ref{fig:pvu:cc} shows an example of connected component with a label-propagation-based method on a given graph. The algorithm follows a synchronous execution model. We find that there may involve two unnecessary vertex updating situations. 1) The vertices have converged before the iteration (e.g. vertex 5 has converged after the first iteration). 2) The vertices are far from the global minimal label but converged with local minimal label (e.g. vertex 6 will get label 1 at the 5th iterations, but with label 2 during the 1st$\sim$4th iteration). Though these updates make no benefit for graph convergence during the iteration, but it is difficult to predict whether update attempt will fail or succeeded in this iteration.

Accurately identifying these unnecessary vertex computations are the key to enhance the GPU processing capability. In the following, we next introduce two efficient yet accurate solutions to predict whether a vertex value would be useful in making a valid vertex update at this iteration towards two situations above.

\subsection{Strong Prediction Condition}
We find that many vertices could be judged as converged by simply according to the vertex value. For example, vertex 5 in figure~\ref{fig:pvu:cc} is converged just after the first iteration, because it has already gotten the smallest label of 1 in the given graph. As the iteration number is increasing, the vertex value changes in a monotonous pattern.
\begin{table}[t]
\centering
  \label{tab:pvu:strong}
  \caption{Strong convergence condition on BFS,WCC and SSSP. $Value[i]$ means the vertex value for various graph algorithms, e.g., the depth for BFS and distance for SSSP. $ccsize[v]$ represents the number of vertices labeled with $v$.}
  \vspace{0.2em}
  \tabcolsep=0.05cm
  \scriptsize
  \begin{tabular}{c|c|c}
  \hline
  {} & condition & definition \\
  \hline
  $BFS$  & $value[i] \le k$ & $k$ is the iteration times\\
  $CC$   & $value[i] \le s$ & $s$ = min\{$v|ccsize[v]^{(k-1)}\neq ccsize[v]^{(k-2)}$\}\\
  $SSSP$ & $value[i] \le l$ & $l$ = min\{$value[v]^{(k-1)}|value[v]^{(k-1)}\neq value[v]^{(k-2)}$\}\\
  \hline
\end{tabular}
\end{table}
We thus list a strongly predictive solution in Table~\ref{tab:pvu:strong}. We prove the condition of convergence separately. And note that the condition is suitable for asynchronous update, naturally suitable for synchronous model.

For BFS, we can conclude that the vertex has converged if the value of vertex i is smaller than the iteration times $k$. It is easy to prove that if there is a shorter unweighted path $r$ ($r < k$ and $r < value[i]$) from the source vertex, the path must have been found in $r$ times iteration. Though the top-down/down-top based BFS also benefits from omitting the update for traversed vertices, but it is not suitable for asynchronous execution model, since the vertex could be traversed with a longer path for the much faster asynchronous update, but not converged actually.

For CC, we can conclude that the vertex has converged if the vertex has been chosen into a converged component. More  specifically, the component size will not increase if the component root (minimal vertex in the component) has packed all the vertex into the component, and we call the component has converged. For a vertex, the optimal situation is finding an component not converged yet. Thus the minimal label could be $s$, the root for the component not converged yet. So the vertex must have converged if the vertex label is smaller than $s$. A special but really useful case is that the vertex has got the minimal label of the whole graph, then can be concluded converged immediately.

For SSSP, we focus on weighted graph without negative edge. We can conclude the vertex has converged if the vertex has found a path shorter than the latest shortest path in the last iteration. It is easy to prove, since the vertex's distance to the source vertex is based on the more closer vertex on the shortest path. It is impossible to get a smaller distance than the closest one at the last iteration.

The strong predictive vertex updating method can judge the vertex converged or not directly from to the vertex value, regardless of the graph topology. And we only need to compared the vertex value with the threshold, then if the vertex value satisfy the condition, the cost for computation and the memory access on this vertex can be all reduced.
\subsection{Weak Prediction Condition}
Though strong prediction can reduce the unnecessary vertex update directly, however, in many cases it is difficult to design a strong converged condition such as data-driven PageRank (PageRank-delta). We thus present an alternative weak predictive technology based on the updating history.

Inspired by using the branch history to predict the branch output~\cite{hennessy2011computer}, we also try to predict the vertex update result with the update history. Firstly we try to find the update patterns for graph processing, so we keep a record of the update history for each vertex with various graph datasets and algorithm (note that we only record the iteration result processed on GPU side with asynchronous execution model). For example, a vertex update history is $[1,1,0,1,0]$ indicating that the vertex value changed at the 1,2,4 iteration, but the update attempt in iteration 3 and 5 failed, indicating the vertex value do not change at these two iterations. And we find an interesting pattern: if the vertex value has changed before, but the value no longer change with the two subsequent iteration, the vertex can be concluded as converged with a high likelihood (91.3\% for SSSP on our dataset, and about 98\% for CC and BFS). We conclude this pattern with a lot of off-line analysis.
%The maximum update gap can be used to predict whether the vertex will be updated or not in this iteration. For example, the vertex value has not changed in the last two iterations. We can also conclude that the maximum gap for most vertices is not greater than 2, indicating that the vertex value would not be changed in this iteration with high possibility.
\begin{figure}
\centering
\includegraphics[scale=0.5]{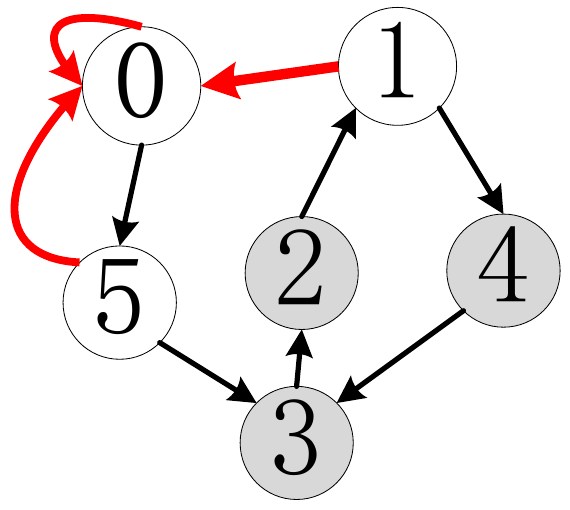}
\vspace{-1em}
\caption{Weakly predictive vertex updating state-transition diagram. We let the vertices with status 2,3,4 to be not updated. If there is a valid update, we return to initial state 0 by following the red arrow, otherwise following the black arrow.}
\vspace{-1em}
\label{fig:pvu:dia}
\end{figure}

Based on the afore-mentioned statistical regularities, we further present a state-transition diagram based solution to guide the vertex updating with low overhead. And the diagram should match the statistical regularities to predict the vertex update result effectively.

Figure~\ref{fig:pvu:dia} shows the state-transition diagram designed for our system. There are six states used in the diagram. The vertices with state 0, 1, 5 indicate that this vertex should make an update attempt during this iteration, and the update attempts for the vertex with state 2, 3, 4 will be canceled during the iteration. At the beginning, all vertices are initialled as the state  0. After the first iteration, the vertex state will get into 5 (dormancy candidate state) if the update attempt failed, or it returns back to 0 if success. The vertices with state  5 will get into 3 (dormancy state) if the update failed again. This design ensures that all vertices with state 3 has failed at least two times and we can assume the vertex has converged with a high likelihood, which matches with the statistical regularities mentioned above. Vertex with state 3 will be dormant and do not make an update attempt at the next two iterations, then memory access and computation can be canceled on the vertices with these status. To avoid the vertex getting into dead state and no longer attempt to update, we add state 1 to make an attempt after the two round of dormancy iterations, but if the update failed again, the vertex will get into state 4 as a punishment.

This method can be integrated into the graph engine naturally. Since Seraph follow a vertex-centric programming model, we just need to allocate a status word for each vertex and initialize it to state 0, then the status can be driven by checking the vertex value changed or not during this iteration, or just turn into the next state when it is in dormancy state. And the state-transition diagram is defined by users as a input for graph engine, making the graph algorithm unmatched with this diagram can also benefit from this method by using other diagram. And though the weak predictive vertex update may be false, we will make an all-pull update to retrieve the active vertices within the dense to sparse switch.

{\bf Discussion}\quad We find that the BFS, CC with strong predictive update method is really efficient, but SSSP can get a better efficiency with weak predictive method. We analyze the update history and find that SSSP with strong predictive vertex update can only reduce a small amount of useless update accurately, but weak predictive update method can remove considerable amount of useless update though may be false with low possibility.

With the predictive vertex updating, many useless update can be identified and further reduce the amount of computation. And these saved computation can be used to squeeze out the real value of loaded data, improve the system efficiency directly.

\section{Evaluation}
In this section, we evaluate the efficiency of Seraph by answering the following four research questions:
\begin{itemize}
    \item {\bf\em RQ1:} How efficient is Seraph compared to existing heterogeneous graph systems? [Section~\ref{sec:evaluation:efficiency}]
    \item {\bf\em RQ2:} How does Seraph scale with varying graph sizes and processing capability? [Section~\ref{sec:evaluation:scalability}]
    \item {\bf\em RQ3:} How effective are pipeline subgraph iteration and predictive vertex updating? [Section~\ref{sec:evaluation:effectiveness}]
    \item{\bf\em RQ4:} How well is Seraph advantageous over other state-of-the-art large-scale graph parallel processing systems? [Section~\ref{sec:evaluation:comparison:other:systems}]
\end{itemize}

\subsection{Experimental Setup}
\label{sec:evaluation:setup}
{\bf Testbed}\quad All tests are performed on a machine where the host side is equipped with an Intel 10-core Xeon E5-2650v3@2.40GHz, 64GB main memory. The GPU accelerator is NVIDIA GTX980 (with 16SMXs, 2048 cores and 4GB GDDR5 memory), which is connected to the host side via PCI Express 3.0 lanes operates at 16x speed. The transmission bandwidth for asynchronous memory copy between CPU and GPU is around 11GB/s.

{\bf Methodology}\quad Graphie~\cite{han2017graphie} and Garaph~\cite{ma2017garaph} are  the most related GPU-accelerated heterogeneous systems to our Seraph. Graphie tracks the active subgraph with a compact data structure, and only transfers the active subgraph from the CPU to GPU to reduce the useless data transfer. It gives the priority to process the subgraph buffered on GPU after last iteration. Garaph dispatches the subgraphs between the CPU and GPU during the dense stage, and processes the subgraphs on the CPU side during the sparse stage.

Unfortunately, neither Graphie nor Garaph is open sourced. For making a comparison with them, we can thus only reference the experimental results reported in their work (just as previous work has also done \cite{han2017graphie,maass2017mosaic}), and evaluate Seraph with the same graph datasets that have been used. Table~\ref{tab:machine:specifications} depicts the detailed machine specification of Seraph against those for Graphie and Garaph. We note that Seraph basically has the worst configuration but gains the best performance as will be discussed in the subsequent experiments.

{\bf Graph Algorithms}\quad We evaluate Seraph using three well-known graph traversal algorithms: 1) {\em Breath-First Search (BFS) } for traversing the graph hop by hop so as to compute the distances of all vertices from a specific vertex ; 2) {\em Connected Components (CC)} for finding a maximal number of subgraphs where any two vertices
can be connected via a chain of paths; and 3) {\em Single-source Shortest Path (SSSP)} for finding a path of a given vertex to every vertex such that the sum of the weights of their constituent edges is minimized. As discussed in Section~4.3, we evaluate BFS and CC with strong predictive vertex updating optimization. SSSP is evaluated with weak predictive vertex updating optimization.

{\bf Graph Datasets}\quad We benchmark the graph algorithms with a variety of graph collections, incuding: 1) 12 real-world graphs (coming from Stanford Large Network Dataset Collection\footnote{http://snap.stanford.edu/data} and Laboratory for Web Algorithmics\footnote{http://law.di.unimi.it/datasets.php}); and 2) 6 large synthesized graphs (generated by the RMAT tool). Table~\ref{tab:graph:collection} depicts the graph collections that we have used for the aforementioned comparison with Graphie and Garaph, and further evaluation.
%\indent And we have tested three typical graph application in our evaluation,including BFS,WCC and SSSP. These application is really general for graph processing and could be the based algorithm for many algorithm. Such as BFS can be the based algorithm for much traversal algorithm. We implement the algorithm in the same manner with Ligra for push and pull efficiency. \\

\subsection{{\em RQ1}: Efficiency}
\label{sec:evaluation:efficiency}
We first evaluate the efficiency of Seraph in comparison to two state-of-the-art GPU-accelerated heterogeneous systems: Graphie and Garaph. In an effort to make a fair comparison, like Graphie and Garaph, our results take both data transfer and kernel execution into account.
\begin{table}
\centering
  \caption{Detailed machine specifications that have been used in Garaph, Graphie and Seraph}
  \label{tab:machine:specifications}
  \tabcolsep=0.05cm
  \scriptsize
  \begin{tabular}{|c|c|c|c|}
  \hline
  Spec. & Graphie & Garaph & Seraph\\
  \hline\hline
  GPU type& NVIDIA Titan Z & NVIDIA GTX1070 & NVIDIA GTX980\\
  On-board memory & 6GB GDDR5 & 8GB GDDR5 & 4GB GDDR5  \\
  Internal bandwidth & 288GB/s & 256GB/s & 224GB/s\\
  CUDA cores & 2688 & 1920 & 2048\\ \hline \hline
  CPU & E7-4830 v3 & E5-2650 v3 & E5-2650 v3\\
  Host memory & 256GB DDR3 & 64GB DDR4 & 64GB DDR3 \\
  %PCI-E & PCI-E 3.0x16 & PCI-E 3.0x16 & PCI-E 3.0x16\\
  \hline
\end{tabular}
\end{table}

\begin{table}[t]
  \caption{Graph datasets used in our evaluation}
   \label{tab:graph:collection}
  \centering
  \begin{threeparttable}
    \tabcolsep=0.08cm
  \scriptsize
  \begin{tabular}{|c|ccc|c|ccc|}
    \cline{2-4}\cline{6-8}
    \multicolumn{1}{c|}{}&dataset&$|V|$&$|E|$&\multicolumn{1}{c|}{} &dataset&$|V|$&$|E|$\\
    \hline\hline
    \parbox[t]{2mm}{\multirow{7}{*}{\rotatebox[origin=c]{90}{Graphie}}}& cage15 & 5.1M & 99.1M& \parbox[t]{2mm}{\multirow{6}{*}{\rotatebox[origin=c]{90}{Garaph}}}&uk2007@1M & 1M & 41.2M\\
    &kron-500 & 2.1M & 182.1M& &uk2014-host & 4.8M & 50.8M\\
    &nlpkkt160 & 8.3M & 221.1M& &enwiki-2013 & 4.2M & 101.3M\\
    &orkut & 3.1M & 117.2M& &gsh-2015-tpd & 30.8M & 602.1M\\
    &uk-2002 & 18.5M & 298.1M& &twitter-2010 & 61.6M & 1,468.4M\\
    &twitter-2010 & 61.6M & 1,468.4M& &sk-2005 & 50.6M & 1,949.4M\\
    \cline{5-8}
    &friendster & 124.8M & 1,806.1M& &RMAT-$k$\tnote{*} & $2^k$ & $2^{k+4}$\\
    \hline
\end{tabular}
\begin{tablenotes}
\item[*] We make $22<k<29$ to create different graph sizes
\end{tablenotes}
\end{threeparttable}
\end{table}

\begin{table}[t]
\centering
  \caption{Seraph vs. Graphie}
  \label{tab:comparison:Graphie}
  \tabcolsep=0.02cm
  \scriptsize
  \begin{tabular}{|c|ccc|ccc|ccc|}
  \hline
  \multirow{2}{*}&
  \multicolumn{3}{c|}{BFS}&
  \multicolumn{3}{c|}{CC}&
  \multicolumn{3}{c|}{SSSP}\\
  \cline{2-10}
  &Graphie&Seraph&speedup&Graphie&Seraph&speedup&Graphie&Seraph&speedup\\
  \hline
  cage15 & 0.63 & 0.095 & 6.63x & 0.23 & 0.066 & 3.48x & 0.24 & 0.54 & 0.44x \\
  kron\_g500 & 0.59 & 0.047 & 12.55x & 0.48 & 0.057 & 8.42x &1.67 &0.135 &12.37x \\
  nlpkkt160 & 6.11 & 0.248 & 24.64x & 1.02 & 0.73 & 1.4x & 7.03 &8.42 & 0.83x \\
  orkut & 0.21 & 0.067 & 3.13x & 0.26 & 0.088  & 2.95x & 0.6 & 0.24 & 2.5x\\
  uk-2002 & 4.3 & 0.23 & 18.7x &5.04 &0.35 & 14.4x & 11.73 &0.85 & 13.8x\\
  twitter & 5.42 & 1.06 & 5.11x & 4.21 & 1.09& 3.86x & 14.67 & 4.8 & 3.06x\\
  friendster & 16.44 & 3.8& 4.32x & 12.46 & 2.3 & 5.42x & 29.24 & 7.07 &4.14x\\
  \hline
\end{tabular}
\end{table}

{\bf Compared with Graphie}\quad Table~\ref{tab:comparison:Graphie} shows the detailed comparative results. It is worth noting that, for Graphie, Garaph and Seraph, {\tt twitter} and {\tt friendster} can not fit into their GPU global memory. Seraph provides a considerable performance benefit with 4.2x on average due to our pipelined subgraph iteration that has less requirement on data transmission.

For the small graphs, all graph data can be fit into the GPU global memory. In this case, though the piplined subgraph iteration might be considered helpless, our predictive vertex updating can still be efficacious in enhancing in-memory computing. As we can see, Seraph significantly outperforms Graphie for almost all cases. For instance, BFS with {\em nlpkkt160} can obtain up to {24.64x} speedup. We should note that, for SSSP with {\tt cage15} and {\tt nlpkkt160} , Seraph has little performance improvement that is because {\tt nlpkkt160} show up meshwork property which is more suitable for a push-based SSSP algorithm~\cite{shun2013ligra}.

%And CGGraph outperform than Graphie in most cases and algorithm,especially for the sssp on cage15 and wcc,sssp on nlpkkt160.  For the small case, all data could be fit into GPU memory and no subgraph need to be transfer to GPU more than once, the efficiency are mainly due to the push-pull optimization and predictive vertex pull optimization. And for the large dataset for twitter and friendster, the speed up are mainly for the stream subgraph schedule method with less I\/O and fully use the global memory to reduce the negative effects for the streaming subgraph processing. \\

\begin{table}
\centering
  \caption{Seraph vs. Garaph}
  \label{tab:comparison:Garaph}
  \tabcolsep=0.05cm
  \scriptsize
  \begin{tabular}{|c|ccc|ccc|}
  \hline
  \multirow{2}{*}&
  \multicolumn{3}{c|}{CC}&
  \multicolumn{3}{c|}{SSSP}\\
  \cline{2-7}
  &Garaph&Seraph&speedup&Garaph&Seraph&speedup\\
  \hline
  uk2007@1M & 0.14  & 0.048& 2.92x & 0.48 & 0.11 & 4.36x \\
  uk2014-host & 0.17 &0.09 &1.89x & 0.57 & 0.236 & 2.42x  \\
  enwiki-2013& 0.28 & 0.05 & 5.6x & 0.7& 0.174 & 4.02x \\
  gsh-2015-tpd & 1.21 & 0.42 & 2.88x & 4.32 & 1.8 & 2.4x\\
  twitter & 3.32 & 1.09 & 3.05x & 12.75 & 4.8 & 2.66x \\
  sk-2005 & 4.47 & 4.01 & 1.12x & 18.13 & 11.13 & 1.65x \\
  \hline
\end{tabular}
\end{table}

{\bf Compared with Garaph}\quad Table~\ref{tab:comparison:Garaph} depicts the comparative results. Since Garaph does not provide the results regarding BFS, we thus test Seraph using CC and SSSP only.
As a consequence, we can see that Seraph has a better performance than Garaph for all graph datasets, with up to 5.6x speedup for in-GPU-memory graphs and 2.6x speedup for out-of-GPU-memory graphs. Note that GTX 1070 used in Garaph has a better configuration than Seraph's, especially with double memory capacity with ours.
There are two reasons for Seraph outperforms Garaph. Though Garaph reduces a part of I/O during the dense stage, but GPU is more powerful than CPU, most subgraphs also have been transferred to the GPU, but Garaph does nothing to cover the gap between computation and data transfer. Second Garaph adopts all-pull method to process the edge during the dense stage, which includes much useless computation, and Seraph reduces these computation with the predictive vertex update. Though Garaph reduces the data transfer during the sparse iteration, Seraph also reduces these IO overhead with a more efficent push-based method.
\begin{figure}[t]
\begin{centering}
%\vspace{-1em}
\subfloat[Out-of-Memory ({\tt uk-2007})]{\begin{centering}
\includegraphics[height=2.9cm,width=3.9cm]{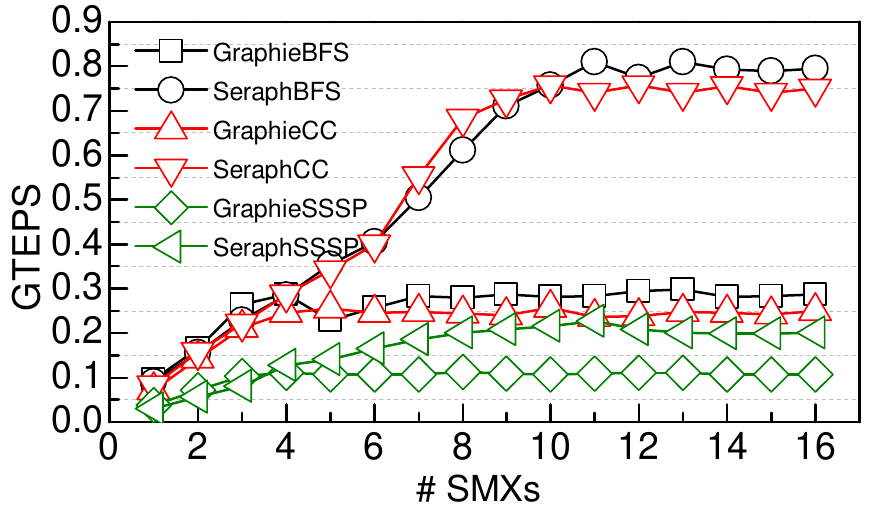}
%\vspace{-0.5em}
\par\end{centering}}
\subfloat[In-Memory ({\tt enwiki-2013})]{\begin{centering}
\includegraphics[height=2.9cm]{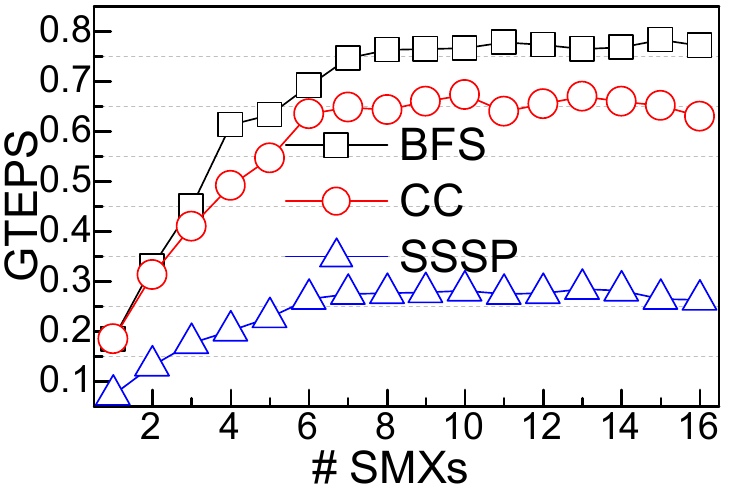}
%\vspace{-0.5em}
\par\end{centering}}
\par\end{centering}
\vspace{-0.5em}
\caption{Performance characterization with a varying number of SMXs. GTEPS represents Giga-scale traversed edges per second.}
%\vspace{-1em}
\label{fig:scalability:scale:up}
\end{figure}

\subsection{{\em RQ2}: Scalability}
\label{sec:evaluation:scalability}
We investigate the scalability of Seraph by: 1) controlling the number of SMXs available for graph processing, and 2) adjusting the graph size that varies from in-memory scale to out-of-memory scale.

{\bf Scalability with varying SMXs}\quad
Figure~\ref{fig:scalability:scale:up}(a) depicts the performance characterization between Graphie's subgraph iteration and ours by benchmarking entire {\tt uk2007} data set (with 105 million vertices and 3.7 billion edges). With our technical highlights of pipelined subgraph iteration and predictive vertex updating, it is revealed that, for all three graph algorithms, Seraph can significantly improve the
performance of graph processing over Graphie as fed with more GPU SMXs. For instance, for BFS, when the number of SMXs reaches at 4, the performance of {Graphie} will be saturated. In contrast, Seraph can continue offering a near-linear performance improvement. CC and SSSP have the similar observation.

We should note that Seraph may also enter into saturation. For BFS, the benefits will stop when the number of SMXs reached at 11. We guess one of underlying reasons may lie in the limited GPU internal bandwidth between SMX and global memory. Figure~\ref{fig:scalability:scale:up}(b) further shows the performance characterization of pure in-memory computing using the small graph {\tt enwiki-2013} that can fit into the global memory. We find that a similar observation has occurred. This yields a conclusion that the latter saturation has nothing to do with the CPU-GPU transmission efficiency. Coping with the internal inefficiency of GPU for graph processing can be interesting future work, which is beyond the scope of this paper.

%the CPU-GPU large scale graph processing is mainly limited by the low data transfer rate, then we put forward two strategy to cover the gap. As Figure 4 shows, without our streaming subgraph strategy and predictive vertex update optimization, firstly the throughput will grow up near-linear because the performance are severly bounded by the computation power. But with the number of SMX increasing, the performance no longer get any benefit from the computation power growing up. We could see that the throughput no longer grow up when the SMX up to 4 for WCC and BFS, and just 3 for SSSP. But with the streaming subgraph optimization and predictive selective scheduling, the throughtput will continue grow up with the number of SMX increasing.  Finally, our system can get around up to 2.66x speed up compared the system without our optimization. One more to say, the throughput for our system could not go up when the SMX number get into a fix number. We attribute to the random memory access for graph processing has used up the global memory bandwidth, and we tested the in-memory graph processing with the limited SMX, we have found this problem also exist in GPU in-memory graph processing.
\begin{figure*}[t]
\begin{centering}
%\vspace{-1em}
\subfloat[BFS]{\begin{centering}
\includegraphics[width=5.2cm]{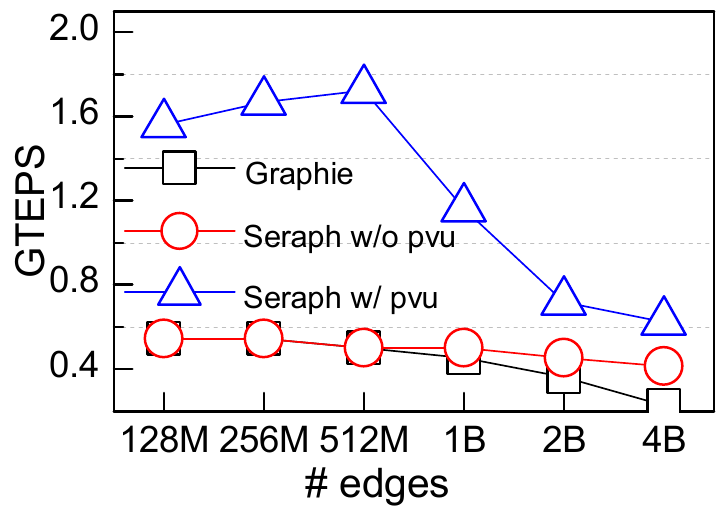}
%\vspace{-0.5em}
\par\end{centering}}
\subfloat[CC]{\begin{centering}
\includegraphics[width=5.2cm]{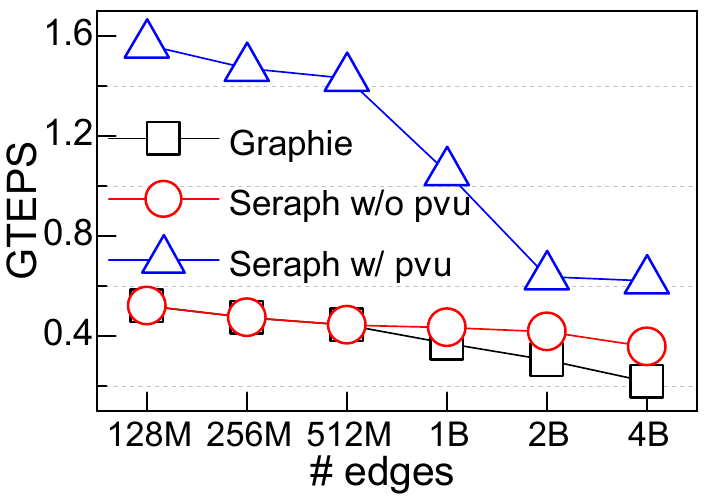}
%\vspace{-0.5em}
\par\end{centering}}
\subfloat[SSSP]{\begin{centering}
\includegraphics[width=5.2cm]{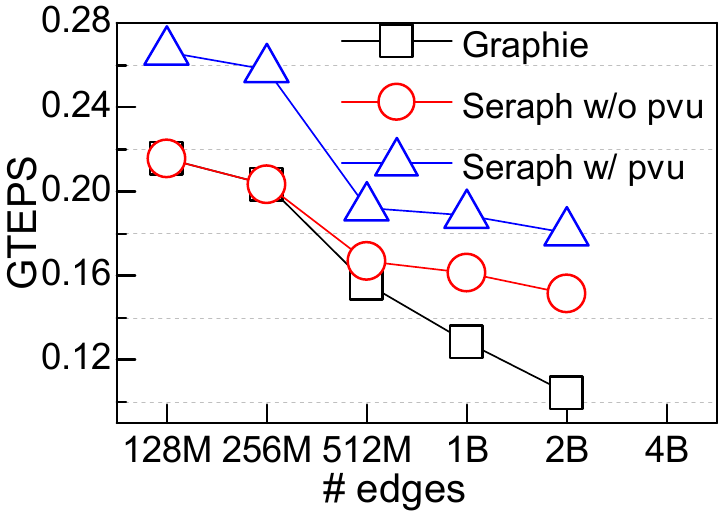}
\vspace{-0.5em}
\par\end{centering}}
\par\end{centering}
\vspace{-0.5em}
\caption{Throughput characterization of Graphie, and Seraph w/ or w/o using predictive vertex updating (pvu) as the graph size of {\tt RMAT} is increasing. SSSP returns nothing for 4-billion edges since it runs out of host memory.}
%\vspace{-1em}
\label{fig:scalability:graph:size}
\end{figure*}

{\bf Scalability with varying graph sizes}\quad
Figure~\ref{fig:scalability:graph:size} depicts the performance characterization using different strategies with different scale of {\tt RAMT} dataset where the edge scale is 16x larger than vertex's. Note that GTX980 has 4G Bytes global memory. BFS and CC works on unweighted graphs, and hence, they can load the unweighted {\tt RMAT} with 1 Billion edges at most into GPU global memory. SSSP works on weighted graph that can have 512 million edges at most into GPU global memory.

%To reduce the impact by the graph property, several synthesis RMAT graph are used in the experiment, we use the ramt generator from ligra with default parameter only with different graph size, the edge number is 16x than the vertex number. The unweighted RMAT23~RMAT25 could fit in GPU memory and unweighted RMAT26-RMAT28 are out of GPU memory, for weighted graph, RMAT23~RMAT24 could fit in GPU memory and larger datasets are out of GPU memory.

Overall, the throughput (i.e., GTEPS) is high when the graphs can be load into global memory. Further, it will go down dramatically when the graph comes to the critical point that is out of global memory because of the CPU-GPU data transmission. To be specific,  predictive vertex updating can introduce more benefits, especially for the in-memory situations. As for large-size graphs, we can still find that our pipelined subgraph iteration and predictive vertex updating contribute to considerable performance benefits.
Note that the host memory will run out for SSSP on {\tt RAMT28} with 4-billion edges, which needs at least 64GB to store the weighted graph data.

%In spite of this, we can still find that our pipelined subgraph iteration and predictive vertex updating contribute to considerable performance benefits. Note that
%The throughput continue to fall since the cache  mechanism make no effect with larger scale graph. With the streaming subgraph optimization, the descend trend slow down obviously, because the streaming subgraph optimization can break the data movement bound and the computation power can be digested again. And RMAT graph get more benefit from predictive vertex update optimization, especially for the in-memory case, for this technology can directly reduce the useless computation. And for large-scale graph, the predictive vertex update can also reduce the computation and fully digest the computation power, but with much useless computation by the multi-iteration, the benefit can not be so obvious.Our system will be out of memory for SSSP on RMAT28, and RMAT28 has 4B edges, which needs at least 64GB to store the weighted graph data. Our system run well for RMAT27, which reflect the memory consume for our system is really good.
\begin{figure}[tb]
\centering
%\vspace{-1em}
\includegraphics[scale=0.65]{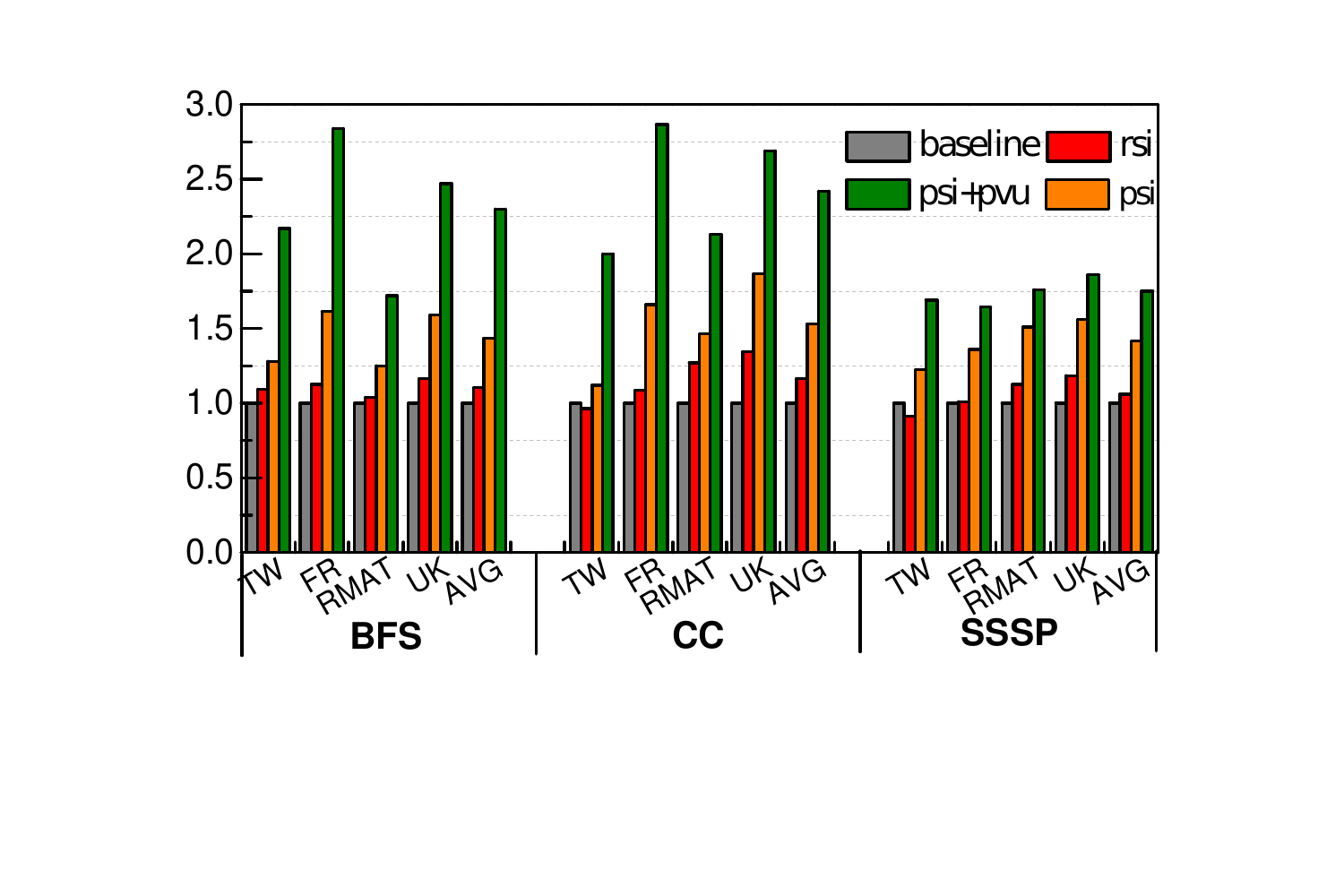}
%\vspace{-1em}
\caption{Performance breakdown and effectiveness evaluation of pipelined subgraph iteration}
%\vspace{-1em}
\label{fig:effectiveness:breakdown}
\end{figure}

\subsection{{\em RQ3}: Effectiveness}
\label{sec:evaluation:effectiveness}
We also evaluate the effectiveness of piplined subgraph iteration (psi) and predictive vertex updating (pvu).

{\bf Breakdown}\quad Figure 9 shows the breakdown results. We conduct the test on four large graph datasets where {\tt TW} is {\tt twitter2010}. RMAT represents RMAT27 and UK indicates the entire dataset of {\tt uk2007}. The baseline is the traditional subgraph processing without multiple iterations~\cite{han2017graphie}.  In comparison to baseline, we can see that psi can offer up to 1.86x speedup for CC on {\tt uk2007}, and especially for large-scale graph,because the subgraph cache optimization is useless when the graph size is much larger than memory size. The predictive vertex update optimization can offer up to 1.79x speed up on twitter, and offer 1.46x speed up on average. The total benefit can get from these two technology is up to 2.86x speed up for CC on friendster, and offer 2.11x speed up on average.

{\bf Effectiveness of pipelined subgraph iteration}
To demonstrate the effectiveness of psi, we also test {naive reentry subgraph iteration (rsi)} that have been used in CLIP~\cite{ai2017squeezing}. For CLIP, we have set the most suitable value of maximum reentry times (MRT) for BFS, CC and SSSP to be 3, 2 and 2, respectively. More details regarding how to set the reasonable value of MRT can be found in~\cite{ai2017squeezing}.

Figure~\ref{fig:effectiveness:breakdown} lists the detailed results.
%release the computation power of GPU with less redundant computation overhead. In this section, we evaluate the benefit achieved by the Piplelined subgraph iteration. \\
%As Figure 8 shows,  We evaluated the subgraph streaming technology efficiency and compared with naive reentry subgraph multiple times used in\cite{ai2017squeezing}. We perform the experiment on four large graph datasets including twitter-2010, friendster, RMAT27 and uk2007, and
We can see that rsi introduces a slight performance improvement in comparison to the baseline, with at most $34\%$ improvement for CC on {\tt uk2007} as an example,and only $10.4\%$ on average. This is because rsi adopts updating the subgraph over and over again, further causing much redundant computation. In comparison to rsi, psi cause less redundant computation due to the pipiline schedule method. As a result, we can get up to 86\% performance improvement for CC on {\tt uk2007}. The averaged performance improvement is by 44.4\%.
%The left one is reuse the cached subgraph but not perform multiple computation on the subgraph, and the middle one is performed multiple computation on the subgraph and cover the data transfer with the data computation, and the computation limited on half global memory. The right one use the streaming evaluation but no selected pull optimization. \\
%\indent Result shows the efficiency for our strategy is obvious. Compared to the subgraph cache strategy, the streaming subgraph strategy can get at most 1.86x speed up, and the speed up is more obvious for large-scale graph. Multi-iteration for subgraph is also effective for large-scale graph, but it much weak than the streaming method. Even more disappointing, the multi-iteration can get no benefit for wcc and sssp on the twitter graph, this is the local converge cause much useless computation but the cache mechanism co/over part gap between computation and data movement.\\

{\bf  Effectiveness of predictive vertex updating}
%predictive vertex updating make benefit for this method could reduce the useless vertex updating according to the vertex value. For the in-memory based graph processing, the efficiency get just from less computation with the. And for the out-of-memory case, with less computation in each iteration, the GPU could perfrom more computation in one data transfer, to sufficiently squeeze out the value of each subgraph.
Relative to strong pvu that can definitely reduce the useless work, the weak pvu has more complex relations. Thus, we choose SSSP to evaluate the effectiveness of pvu for better demonstration.
%For further analyze the efficiency for weak predictive vertex updating optimization, we make a trace to record the vertex status during each iteration for sssp on cage15.

\begin{figure}[t]
\begin{centering}
%\vspace{-1em}
\subfloat[Status]{\begin{centering}
\includegraphics[width=3.9cm]{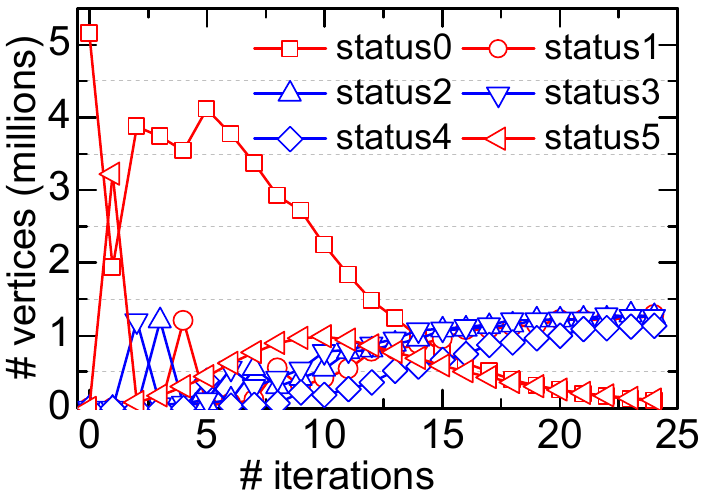}
%\vspace{-0.5em}
\par\end{centering}}
\subfloat[Computation amount]{\begin{centering}
\includegraphics[width=3.8cm]{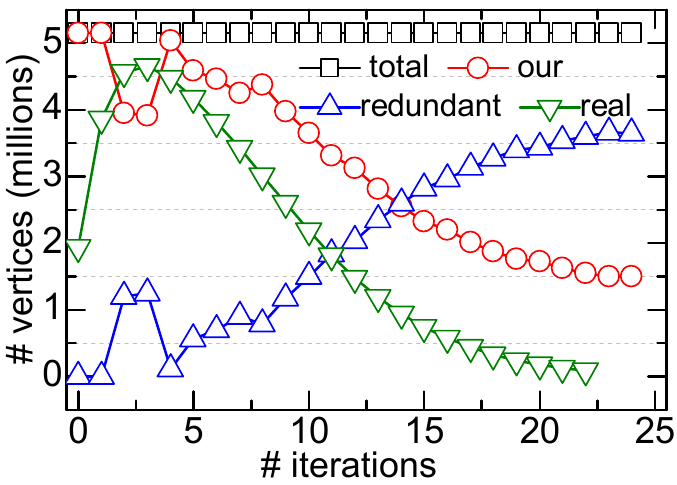}
%\vspace{-0.5em}
\par\end{centering}}
\par\end{centering}
\vspace{-1em}
\caption{Effectiveness evaluation of predictive vertex updating using SSSP with {\tt cage15}. (a) The variation in the number of vertices for each status; (b) The variation in the amount of computation.}
\vspace{-1em}
\label{fig:effectiveness:pvu}
\end{figure}

Figure~\ref{fig:effectiveness:pvu}(a) counts the number of vertices for each status during each iteration. It shows that the number of vertices that involve a data update (i.e., with status 0, 1 and 5) is significantly reduced as iteration goes deeper. That is, there is a large amount of redundant computation (i.e., with status 2, 3 and 4) that does not lead to a valid vertex updating is increasing during each iteration. Figure~\ref{fig:effectiveness:pvu} further sums up the number of vertices (with status 0/1/5) as {\tt our} label, and those (with 2/3/5) as as {\tt redundant} label. {\tt Real} indicates the number of vertices that involve the valid updating during each iteration. Compared to the total amount of computation in prior work~\cite{shun2013ligra, ma2017garaph}, our optimization can basically reflect the real situation, and also, dramatically reduce their redundant computation for performance speedup.

%The figure 9(a) count the status number for vertex on each iteration, and result shows the number of vertex need to be updated(status 0,1,5) go down with the iteration number increasing, and the number of vertex get into dormant state(status 2,3,4) go up with the iteration number increasing. We sum up the vertex get into dormant state(status 2,3,4) and show the result in figure 9(b), the horizontal line reflect the number of vertex to be update during each iteration(just all vertex) and the real line reflect the valid updating during each iteration. Result shows our optimization can get a close updating mechanism to be real line, then reduce the useless computation. \\

\begin{figure}[tb]
\centering
%\vspace{-1em}
\includegraphics[scale=0.8]{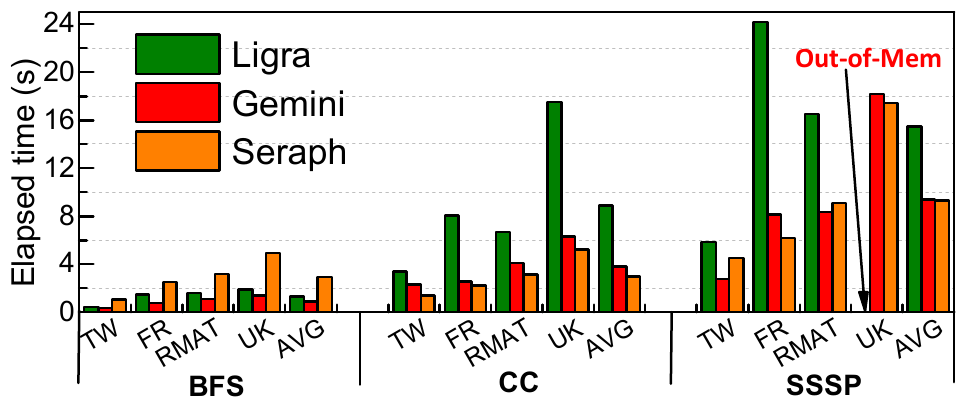}
\vspace{-1em}
\caption{Performance comparison with other state-of-the-art solutions for large-scale graph processing}
\vspace{-1em}
\label{fig:evaluation:other:systems}
\end{figure}

\subsection{{\em RQ4}: vs. Other Advanced Large Scale Graph Processing Solutions}
\label{sec:evaluation:comparison:other:systems}
We finally compare Seraph with other state-of-the-art solutions for large-scale graph processing: 1) Ligra~\cite{shun2013ligra}, a host-memory based solution; and 2) Gemini~\cite{zhu2016gemini}, a distributed solution. Since Ligra is memory consuming~\cite{ma2017garaph}, we thus extend to use a node with 256GB memory for testing it. Gemini is set on a 4-node cluster. Each node has the same configuration as the one for Ligra.

Figure~\ref{fig:evaluation:other:systems} depicts the results. In comparison to Ligra, it can be seen that Seraph is much efficient for CC and SSSP with up to 3.9x speedup. Heterogeneous solution with an integration of specialized accelerator can provide a great potential over traditional shared memory solution. Even compared to Gemini (with more computing and storage resources), Seraph can also obtain the comparable results. For instance, our results for CC on all datasets are superior to the ones using Gemini, with up to 1.64x speedup. Though Gemini is a good scale-out solution, heterogeneous solution can also offer comparable efficiency, mainly thanks to less network communication and fewer redundant computations.

Note that both Ligra and Gemini show a better efficiency than Seraph for BFS. The underlying reason lies in some specialized optimization they have used. To be specific, vertex update will break early during the dense iteration if the vertex has found a neighbor vertex has been traversed.So the computation for BFS is much sparse than CC and SSSP,  We should also note that such tailor-made optimization can be also implemented into Seraph framework for accelerating BFS.
\vspace{-1em}
\section{Related Work}
{\bf Heterogeneous Graph Systems}\quad
%Accelerator based graph system can provide much computation power in one node. Limited by the pitiful global memory size, these system could not process large-scale graph exceeding the GPU memory. Totem first distribute the graph among CPU and GPU memory with a hybrid graph partition to process large-scale graph on heterogeneous platform, and \cite{} stream the subgraph from host memory to GPU memory to process more tasks on GPU. Graphie try to improve the coalesced memory access efficiency with a novel rename method, and Garaph reduce the write conflict through replication-based gather. However, our experiment shows the GPU computation engine is sufficient enough but the system performance is severely bounded by the low data movement. But neither Graphie or Garaph could break the gap and release the true computation power of GPU.\\
Heterogeneous architecture has integrated the advantageous resources of different devices for satisfying different demands of graph processing. A wide spectrum of efforts have been put into developing specialized graph accelerators via single-~\cite{Zhang:2017:BPF, Zhou:2016:HTE, Jin:2017:ICDCS} or multi-FPGAs~\cite{Dai:2017:FEL} for energy-efficiency purposes. There also emerge a number of GPU-accelerated heterogeneous graph systems~\cite{wang2016gunrock,hong2017multigraph,liu2015enterprise,khorasani2014cusha} for supporting high-performance large-scale graph processing.
%GPU-Accelerated graph engine\cite{wang2016gunrock,hong2017multigraph,liu2015enterprise,khorasani2014cusha} mainly focus on improving the coalesced memory access efficiency or load balance among thousands of threads. However, all-in-GPU memory is unpractical for the large-scale graph.
Graphie~\cite{han2017graphie} transfers the active subgraph to GPU and reuses the cached subgraph in the next iteration to reduce the I/O amount. Garaph \cite{ma2017garaph}
streams edge data asynchronously to the GPU for graph processing.

As discussed before, these prior graph systems may still fall into limited performance when handling large-scale graphs due to their inefficient data transmission. This paper first (to our best knowledge) proposes to leverage multi-time subgraph iteration to break through this limitation, enabling the scale-up efficiency of heterogeneous graph systems.

{\bf Distributed Graph Systems}\quad
A large amount of research seeks help from distributed deployment that can aggregate more resources than single machine for processing large-scale graphs. The primary task of distributed graph systems is to obtain well-cut graph partitions~\cite{gonzalez2012powergraph,chen2015powerlyra,gonzalez2014graphx,avery2011giraph,zhu2016gemini} so as to minimize the communication across machines. A few recent studies use the emerging high-speed network (e.g RDMA)  to reduce the communication overhead~\cite{wu2015g, Shi:2016:FCR}.
Gemini presents a series of adaptive runtime optimizations with sparse-dense switching, locality-aware and NUMA-aware features, enabling an attractive scale-out efficiency.

In comparasion to these distributed designs, it is verified in our evaluation (Section~7.5) that heterogeneous solutions, in spite of involving fewer resources, can be also or even more promising in practice for large-scale graph processing because of less communication cost and fewer redundant computations.

{\bf Disk-based Graph Systems}\quad There also exist many disk-based systems to support large-scale graph processing. GraphChi~\cite{kyrola2012graphchi} proposes parallel sliding windows to lead only non-sequential accesses to the disk. GridGraph~\cite{zhu2015gridgraph} uses 2-level hierarchical partitioning to reduce the I/O amount. TurboGraph~\cite{han2013turbograph} presents a pin-and-slide model to fully exploit the multicore and I/O parallism.
Since the significantly low disk-to-memory bandwidth, disk-based graph systems are orders-of-magnitude slower than heterogeneous solutions.

{\bf In-Memory Graph Systems}\quad For
in-memory graph processing, the graph data only needs to be copied at the beginning. Once data is ready, the processors can process the graph without any data transmission until finished~\cite{malicevic2017everything,shun2013ligra,nguyen2013lightweight}. Usually, in-memory graph systems, such as Gunrock~\cite{wang2016gunrock}, can provide orders of magnitude performance improvement over heterogeneous implementations. However, limited to global memory, existing dedicated accelerators (e.g., GPU) can not store real-world graphs (with more than billions edges)~\cite{zhang2015numa, shun2013ligra}. Considering the inefficient data transmission, the potential of accelerators is also significantly underutilized. On the contrary, combined with pipelined subgraph iteration, we further present predictive vertex updating to better exploit the GPU processing capability for enhancing the performance of heterogeneous graph systems.

\section{Conclusion}
\vspace{-1em}
\indent There remains tremendously challenging for scaling up the performance of heterogeneous graph systems due to the well-known interconnect transmission inefficiency. To cope with problem, by iterating each subgraph multiple times, it is observed that the heterogeneous graph systems can further obtain an improvable performance by enhancing GPU processing capacity.

With this guideline, we develop Seraph integrated with two technical innovations. First, we present a pipelined subgraph iteration to maximize the information propagation of each subgraph to other ones for fully exhausting its value. Second, we propose two efficient vertex updating solutions to predict unnecessary vertex computations for better supporting pipelined subgraph iteration. Our results demonstrate that Seraph outperform the state-of-the-art heterogeneous graph systems Graphie and Garaph by 5.42x and 3.05x, respectively. Seraph can be also scaled up over existing heterogeneous graph system. Our comparative results also reveal that Seraph can achieve impressive performance in comparison to state-of-the-art CPU-based (i.e., Ligra) and distributed graph pro-
cessing systems (i.e., Gemini).

{
\bibliographystyle{IEEEtran}
\bibliography{sample-bibliography}
}
% that's all folks
\end{document}